\documentclass{he_symp}
\usepackage{psfig,graphicx,epsfig}
\usepackage{color}
\usepackage{amsmath,amssymb,epic,eepic,array}
\begin{document}
\renewcommand{\FirstPageOfPaper }{ 13}\renewcommand{\LastPageOfPaper }{ 25}

\title{X-rays from Supernova Remnants}
\author{B. Aschenbach}
\institute{Max--Planck--Institut f\"ur extraterrestrische Physik, Giessenbachstra{\ss}e, 85740 Garching, Germany}
\maketitle

\begin{abstract}
A summary of X-ray observations of supernova remnants
is presented including the explosion fragment A of the Vela SNR, Tycho, N132D,
RX J0852-4622, the Crab Nebula and the 'bulls eye', and SN 1987A, high-lighting the progress made
with Chandra and XMM-Newton and touching upon the questions which arise
from these observations and which might inspire future research.
\end{abstract}

\section{Introduction}

X-ray research of supernova remnants (SNRs) in the 1990's was
dominated by ROSAT and ASCA. Whereas ROSAT put together
the first telescopic survey of our Galaxy and produced well resolved
images and coarse spectra, ASCA added high quality spectra and 
extended the spectral range up to 7 keV and more. 
New scientific grounds were broken on the propagation  of SNRs
into the interstellar medium (ISM), the physics of  supernova explosions
by the discovery of fragments of the progenitor stars which shed
new light on the question of element mixing in supernova explosions, 
and on the production of tens of
TeV electrons in SNR shells, reviving the discussion of the origin
of cosmic rays. Present research is led by the discoveries made with 
Chandra and XMM-Newton. They produce excellent images and spectra
of unprecedented spatial and spectral resolution over a spectral
bandpass never accomplished before. Images of various SNRs
 seem to show
 a distinction between the matter shocked by either the blast
wave rushing into the ISM, more precise the ambient medium, or the 
reverse shock heating the progenitor's ejecta, which includes 
differences in elemental
abundances and their spatial distribution. A limited selection 
of new results  will be presented. Already now we see new scientific
challenges arising which are subject for future research beyond
Chandra and XMM-Newton but not only for future X-ray missions but
for radio and optical observations as well.

\section{Compilation of observational results and associated questions}

\subsection{The explosion fragments of the Vela supernova remnant}
A couple (7) of extended X-ray emission regions outside of the general boundary 
of the Vela SNR have been discovered in the ROSAT all-sky survey, which were 
suggested to be composed mainly of stellar fragments
of the progenitor star (Fig. 1, Aschenbach et al., 1995). 

\begin{figure}
\centerline{\psfig{file=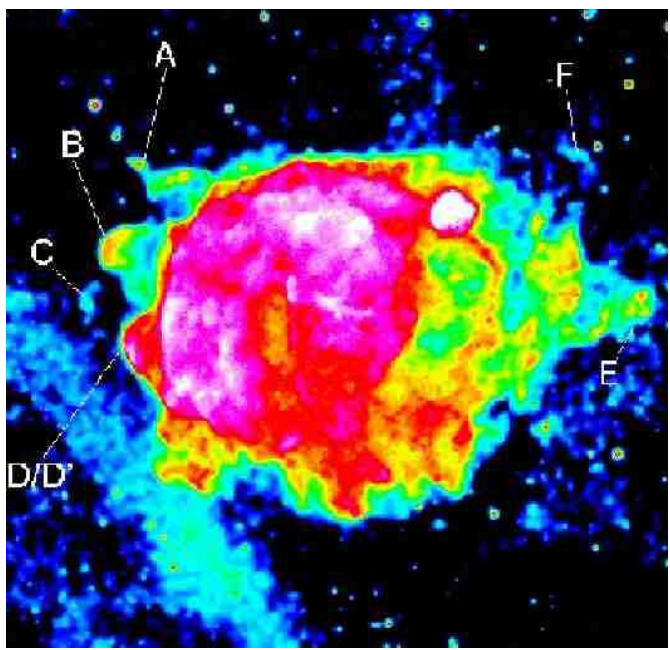,width=8.8cm,clip=} }
\caption{ROSAT all-sky survey image of the Vela SNR and the fragments A -- F 
(Aschenbach et al., 1995).
\label{image}}
\end{figure}

The spectrum of fragment A measured with ASCA 
indicates a significant overabundance
of Si with respect to all other lighter elements (Tsunemi et al., 1999) 
 which is consistent with the view that this fragment
originated deep in the interior of the star, but a quantitative analysis 
was difficult to do.
Fig. 2  shows the XMM-Newton EPIC pn-camera image of fragment A, of which 
the head, 10$\arcmin$ x 5$\arcmin$ in size,
is the brightest part. Both the tail, which extends back to the SNR 
boundary, and the head show
pronounced structure of surface brightness. Fig. 3 is a display 
of the XMM-Newton EPIC spectra, i.e. the pn-camera spectrum and the MOS spectra, of the
total fragment which reveal the presence of strong Mg (1.38 keV) and Si (1.86 keV) lines but a 
significant deficit of high ionization emission lines from lighter elements. Nonetheless, 
weak emission lines from  the light elements are present best visible in the MOS spectra: 
lines from CVI at 0.361 keV and 0.435 keV 
(seen for the first time from an SNR), NVI and  
lines from OVII and OVIII, which all are important to fix the abundance of an element to that of oxygen, 
for instance, the ratio of which is fairly insensitive to spectral models.   
The spectral analysis demonstrates that the Si/O abundance ratio in
the head region is about ten times solar, and that of Mg/O about 4 times solar.
This result confirms that fragment A is associated with the ejecta (Aschenbach \& Miyata, 2002). 
It is noted that there is no 
indication of any sulfur H- or He-like emission lines. The upper limit of the Si/S abundance 
ratio is just consistent with current core collapse supernova models. The definite 
determination of this ratio could, however,
be a crucial test. Observation times of $> 100$ ks with XMM-Newton would be needed. 

The tip of the head of fragment A is offset from the Vela SNR explosion center by 5.3$\sp{\circ}$, which 
implies a  mean proper motion of 1.7"/yr for an age of $\tau$ = 11.4 kyr, which is the age estimated from 
the spin down rate of the Vela pulsar, or more precise the value of $\frac{P}{2~\dot{P}}$ with P the current rotation 
period and $\dot{P}$ its time derivative. Comparison of the ROSAT, Chandra and XMM-Newton images might reveal the 
current proper motion and whether fragment A 
 has slowed down since its creation. The mean true velocity v is given by the 
proper motion times the distance and amounts to v = 2000 km/s~$\times$~d$\sb{250}/\tau\sb{11.4}$ for d = 250 pc.  
v can be compared with model calculations for core-collapse SNe, like the ones of 
Shigeyama et al. (1996), who present element resolved mass profiles and velocity profiles for type II-P 
and II-b SNe. For a 20 M$\sb{\odot}$ progenitor without mixing of the Si/Ni$\sp{56}$ core 
into the upper layers most of the Si/Ni$\sp{56}$ is expelled at velocities less than 900 km/s, 
whereas in the 20 M$\sb{\odot}$ progenitor model with mixing the bulk of the Si/Ni$\sp{56}$ mass  
moves at $\sim$1400 km/s. If correct, in either case a reduction of d$\sb{250}/\tau\sb{11.4}$ 
is required by about a factor of 1.5 to 2 or even more if the fragment has decelerated 
significantly. Since d = 250 pc appears to be a lower limit $\tau$ should be increased which 
would put the age in the range of at least 17 to 23 kyr, an age favored by Aschenbach et al. 
(1995) and independently by Lyne et al. (1996) by their analysis of the time dependence 
of the pulsar rotation rate. One of the key questions is, whether pulsars are borne with a rotation 
rate much faster than the present one. If not, $\frac{P}{2~\dot{P}}$ is not a reliable measure of the true age.  
      
Obviously, massive sub-stellar, non-core fragments are created in
 core-collapse SNe, but the
immediate question is how such an explosion fragment can survive over a distance of $\sim$20 pc
for $>$11 kyr as in the case for the Vela SNR without being dissolved. 
Most recent 2-D hydrodynamical codes including time dependent ionization and 
radiation have
been used to follow the
fate of such a fragment (Brinkmann, W., 2002). The first results 
show that the fragment can
survive if of sufficiently high Mach number and, more importantly, if the
initial matter density of the clump exceeds the ambient matter density
by a significant amount. Fig. 4 shows the temperature distribution 
(10$\sp 6$ K to a few 10$\sp 7$ K) of such a
clump of matter; note that the x-axis is highly non-linear in scale expanding 
along x with increasing distance from the central symmetry line.
Interestingly, such conditions would favor the survival of high density
structures from deep inside the progenitor and rather diffusive large-scale
structures are expected to be left from the top layer matter. In that respect 
X-ray spectral measurements of fragment E (c.f. Fig. 1), which are coming up with XMM-Newton, 
 are interesting.

\begin{figure}
\centerline{\psfig{file=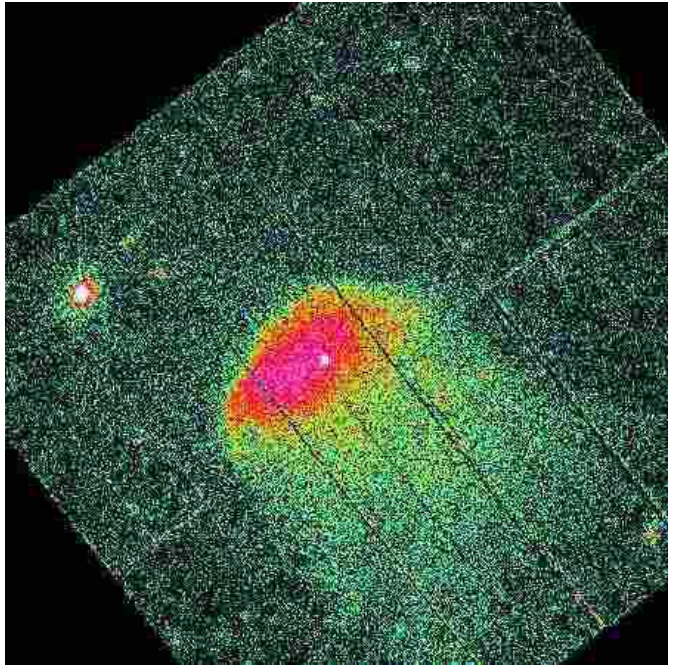,width=8.8cm,clip=} }
\caption{XMM-Newton EPIC pn-camera unsmoothed image of the Vela SNR fragment A.
\label{image}}
\end{figure}

\begin{figure}
\centerline{\psfig{file=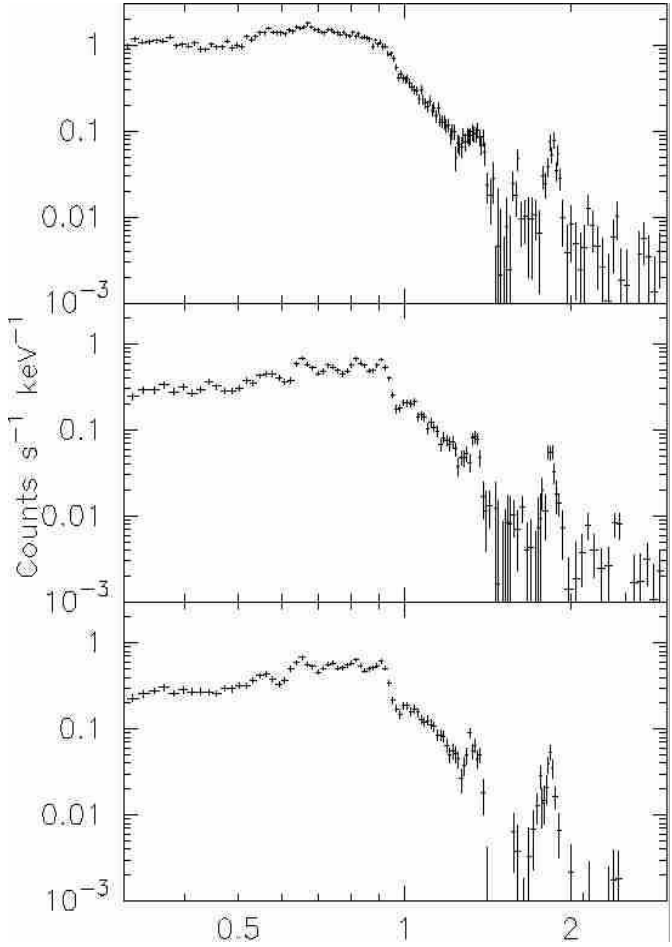,width=8.8cm,height=12.5cm,angle=0,clip=} }
\caption{XMM-Newton EPIC spectra of the Vela SNR fragment A; from top to bottom 
 pn spectrum, MOS1 spectrum, MOS2 spectrum.
\label{image}}
\end{figure}

\begin{figure}
\centerline{\psfig{file=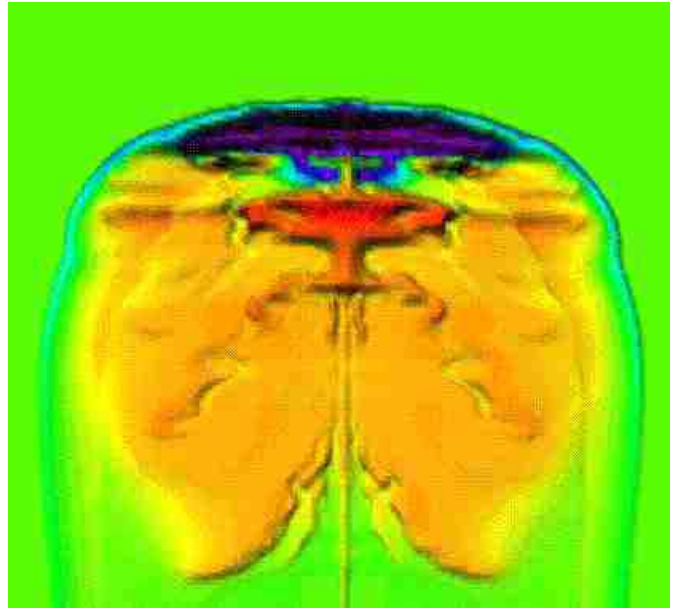,width=8.8cm,angle=0,clip=} }
\caption{Temperature distribution of a supersonic clump of matter
from 2-dimensional hydrodynamical calculations (Brinkmann et al., 2002). 
\label{image}}
\end{figure}


\subsection{The Tycho SNR}

The remnant of the supernova observed by Tycho Brahe in 1572 is an excellent target of research
concerning the physics of the explosion and expansion of a type Ia supernova for which Tycho has
been considered for a long time to be the proto-type.
Type Ia supernovae are believed to originate from the deflagration or detonation  of an
accreting white dwarf most likely living in a binary system.
X-rays from Tycho were recorded in the pioneer days of X-ray astronomy, which already
revealed the presence of high temperature plasma and a bright emission line originating from 
the Fe-K$\sb{\alpha}$ transition. High resolution X-ray images were obtained with the $\it{Einstein}$ and
later
with the ROSAT HRI.  Fig. 5 shows the image of the Tycho SNR
obtained with the XMM-Newton MOS1 camera. The remnant has an almost perfect circular appearance with 
a low surface brightness ring enclosing most of the emission. 
This appearance has brought forward the idea that the outer annulus is
associated with the blast wave shock and that the knotty ring further inside 
contains the reverse shock heated
ejecta. Previous X-ray spectra, compiled from the entire remnant, have shown the presence of many H- and He-like K 
emission lines up to
the Fe and Ni K-lines, including a blend of Fe-L lines. With XMM-Newton spatially resolved spectra
have become available. They show a highly non-uniform distribution. For instance, the south-eastern rim, which
appears to lead the remnant's expansion there, is dominated by three knots. The spectra of each of the
knots can be fitted with the same temperature and the same ionization timescale but the abundances
of Si, S and Fe differ remarkably (Decourchelle et al., 2001). The most northern knot shows the highest abundances
of S and Si, apparently a few times solar,  and some Fe, whereas
the most southern knot basically contains no Fe at all demonstrated
by the absence of both the Fe-K and Fe-L lines. In comparison with the other two knots a factor of
10 less Fe seems to have been mixed into the Si layer at the explosion when the knots formed.
The enhancement of Si and S indicates that the matter is from the deeper ejecta and the abundance
variations point to a mixing of the deeper Fe layer changing from place to place. In that sense these
knots resemble the fragments observed in the Vela SNR, except that the Tycho knots are much younger and
may reflect more directly the explosion physics. The angular separation of the knots along the 
circumference is
 $\sim$15$\sp{\circ}$ and the linear extent of each knot is $\sim$0.3 pc, which is remarkably
close to the elongation of fragment A of the Vela SNR despite their different ages.

On a larger scale the XMM-Newton spectra also provide 
a much better test bed for
the
predictions of the deflagration models developed for type Ia SNe.  
 Images taken just in the light of the K emission line of one of the elements
demonstrate a generally rather regular stratification such that the lighter
elements are found predominantly at large radii and vice versa with the
'Fe ring' having the smallest radius (Fig. 6). Since kinematics studies based on a comparison of the early 
$\it{Einstein}$ image, the ROSAT images and the recent XMM-Newton image show that
Tycho is expanding still almost undecelerated the images also reflect the velocity
distribution of the ejecta. Fig. 6 confirms the onion shell model with the
dominant elements ordered in shells following their atomic number (Stadlbauer \& Aschenbach, 2001). Also, the
ejection velocity at the explosion follows a homologous sequence. This
 was exactly what was predicted by the
deflagration models in the early 80's but one should keep in mind that these
models were one dimensional and did not allow for radial convection, mixing
and instabilities. But it seems that the average ejection velocity of each
element was predicted close to correct.

Fig. 7 shows three individual spectra from a radial, 12.5$\sp{\circ}$ wide
cut through the south-west of the remnant together with the brightness
profile. The Fe-K line shows up only at small radii and is absent
even at positions where the brightness peaks. In this area lines from Si and S
are most prominent. The emission from these areas is clearly associated
with shocked ejecta. Going further outside the line emission decreases but
the Si and S lines are still present. Given the increase in temperature
towards the edge the abundances of Si and S are excessively high by
a few times the solar value. This is not at all expected from a region
the emission of which should be dominated by blast wave heated interstellar
matter, because the abundances should be close to solar. This discrepancy
can be explained by that we either do not see the blast wave front at all
or there exist high speed precursors of the Si/S shells. The difficulty to verify 
the one or other option is the fact that Tycho as most of the young SNRs show 
a high energy continuum which is evident from the spectrum beyond the Fe-L lines. 
In case of Tycho the spectrum of the outer annulus requires a significant 
contribution from a power law or bremsstrahlung component, which in the latter 
case is associated with an emission measure exceeding that of the lower temperature 
components by a large factor. If the bremsstrahlung option is the correct one 
the abundance values of Si and S are highly unreliable. A detailed, spatial resolved 
spectral analysis of the outer annulus along its circumference might shed more light 
on this issue. Imaging observations at even higher energies beyond 10 keV will help as well. 

\begin{figure}
\centerline{\psfig{file=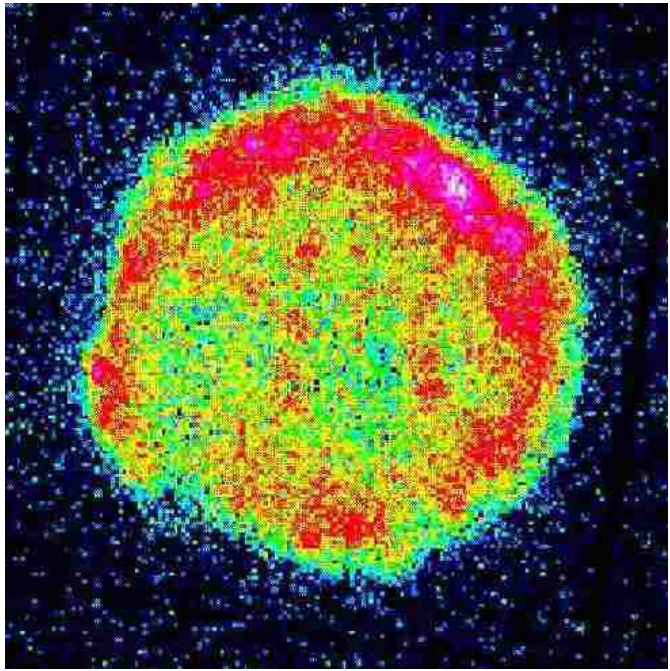,width=8.8cm,clip=} }
\caption{XMM-Newton EPIC MOS1-camera unsmoothed image of the Tycho SNR.
\label{image}}
\end{figure}

\begin{figure}
\centerline{\psfig{file=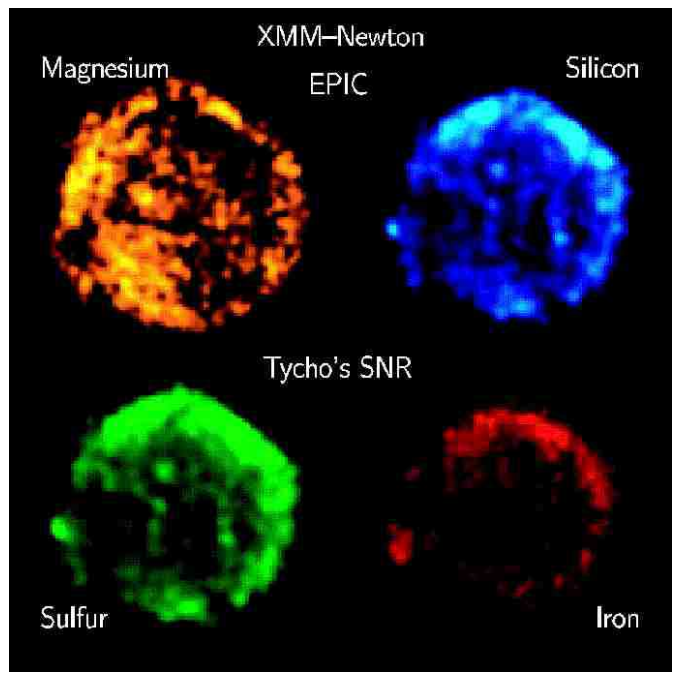,width=8.8cm,clip=} }
\caption{The XMM-Newton EPIC pn-camera images of the
Tycho SNR resolved in the emission lines of Mg, Si, S, Fe.
Sub-images are at the same scale.
\label{image}}
\end{figure}


\begin{figure*}
\centerline{\psfig{file=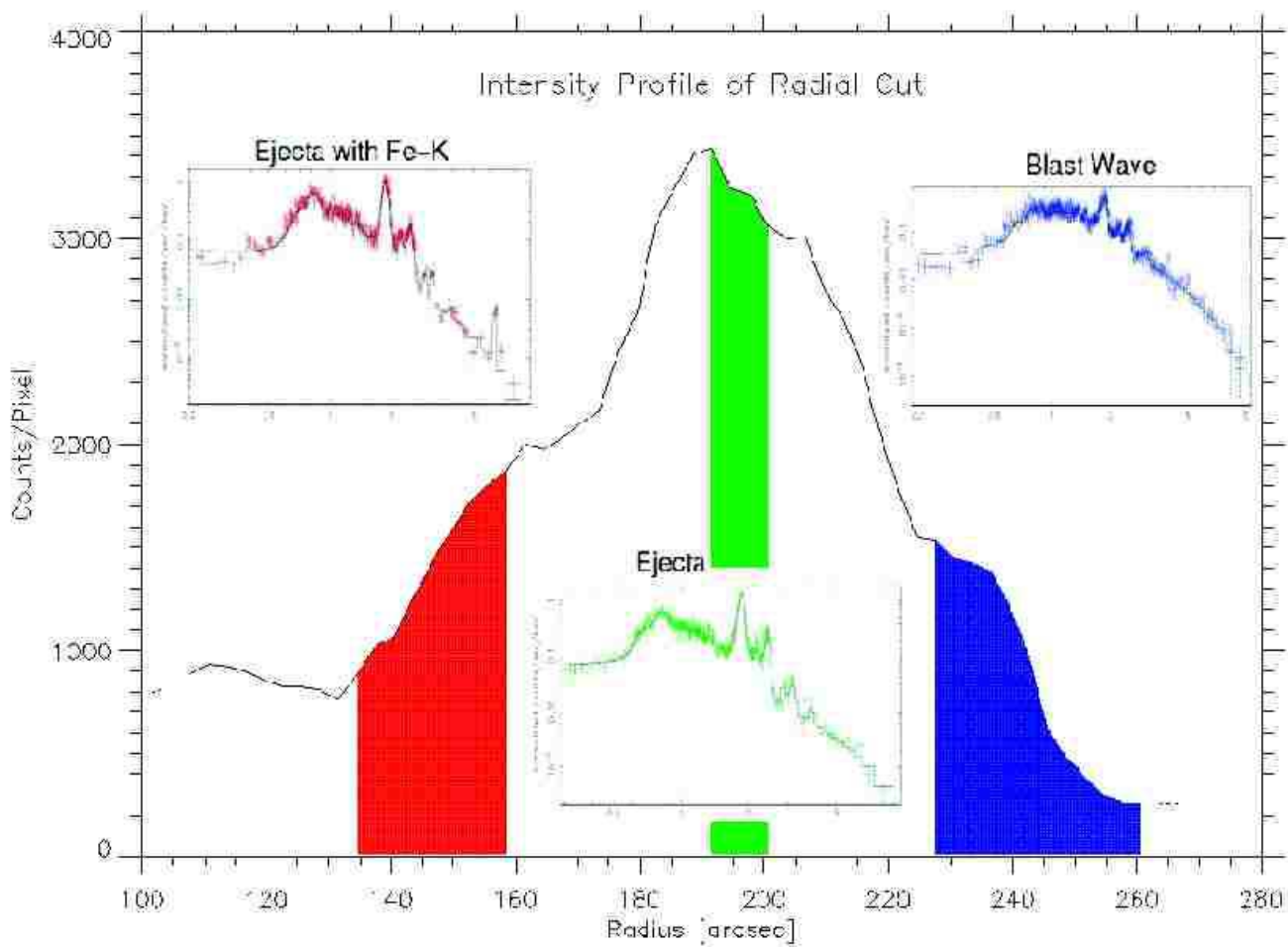,width=17.5cm,clip=} }
\caption{Three EPIC pn-camera spectra of a narrow cut through
the south-west of the Tycho supernova remnant (Stadlbauer \& Aschenbach, 2001).
\label{image}}
\end{figure*}


\subsection{N132D in the Large Magellanic Cloud}

N132D, a supernova remnant in the LMC is estimated to be $\sim$3000 years old. The remnant, which resembles
the Cygnus Loop in its general surface brightness appearance with intriguing filamentary structure
and an apparent blow-out to the north-east, has been imaged by Chandra in great detail.
Fig. 8 shows the image obtained with the XMM-Newton EPIC pn-camera. The break-out and the 
filamentary structure 
of the limb as well as a bright filament running across the middle of the image are clearly visible. 
Superimposed is the first
high quality broad-band X-ray spectrum of the entire remnant obtained with the XMM-Newton EPIC-pn camera.
Emission lines of Si, S, Ar, Ca and Fe are very prominent. By itself this spectrum is amazing
because it shows the capability of XMM-Newton: the spectrum covers a flux range of 4.5 orders of
magnitude in one exposure. Fig. 9 shows the high-resolution spectrum obtained with the XMM-Newton
RGS spectrometer, through which 31 emission lines could be identified (Behar et al., 2001). For the first time
lines from H-like N ($\# 24$) and C ($\# 31$) are seen and they might provide a link to UV observations.
Modeling of the RGS spectrum indicates a temperature of $\sim$0.6 keV with a definite upper limit
of 1 keV. With this temperature and the abundance of Fe apparent in the RGS spectrum in ionic states
from FeXVII to FeXXI essentially no
emission of  Fe-K is expected, contrary of what the EPIC-pn spectrum shows (c.f. Fig. 8).
There must be a high temperature component which produces the high energy continuum and the
Fe-K line emission. Interestingly, progressively higher ionic charges of Fe up to FeXXIV
are not detected although within the energy range of RGS (strong lines expected at
10.6 and 11.2 \AA ), which means that
plasma of any intermediate temperature between 0.6 keV and a few keV is largely absent.
Furthermore, the spatial distributions of the low temperature and high temperature
components are totally different. Whereas the images taken at low ionic charge follow the  
filamentary structure (c.f. Fig. 6, Behar et al., 2001), although different in detail 
for instance in OVII and OVIII, the high energy
continuum and the Fe-K line distributions are fairly uniform across the remnant (Fig. 10). There appears
to be a volume filling high temperature plasma whereas the low temperature plasma is in filaments. 
What causes the temperature stratification with a gap in temperature between 0.6 keV and 2 keV at least?
Why are the low Z and high Z but low ionization ions found predominantly found in filaments whereas 
the high ionization Fe ions and the high energy continuum distributed fairly uniform across 
the remnant? 

The spatial distribution of the high energy continuum is not unique to N132D, a similarly uniform
distribution has also been found for Tycho and Cas A (Bleeker et al., 2001, Willingale et al., 2002). 
It has been suggested that the high energy
continuum is at least
partly due to synchrotron radiation from highly relativistic electrons ($\sim$10's of TeV)
accelerated by diffusive shocks. Diffusive shock acceleration should dominate at the outer shock 
and the emission should be prominent 
there which is observed neither for Cas A nor for N132D nor for Tycho. If the high energy continuum 
is actually due to synchrotron radiation
the acceleration mechanism is more likely to be of turbulent nature acting throughout the remnant 
but on a local scale.

\begin{figure}
\centerline{\psfig{file=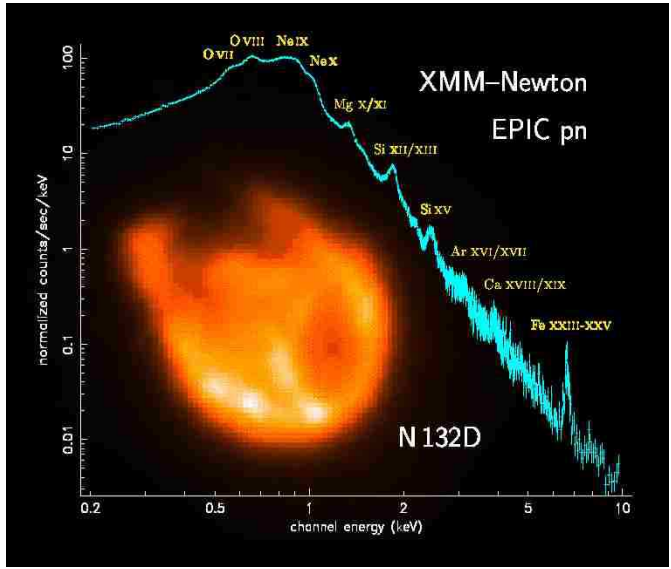,width=8.8cm,angle=0,clip=}}
\caption{XMM-Newton EPIC pn-camera image and total spectrum of the SNR N132D in the Large 
 Magellanic Cloud (Behar et al., 2001).
\label{image}}
\end{figure}

\begin{figure}
\centerline{\psfig{file=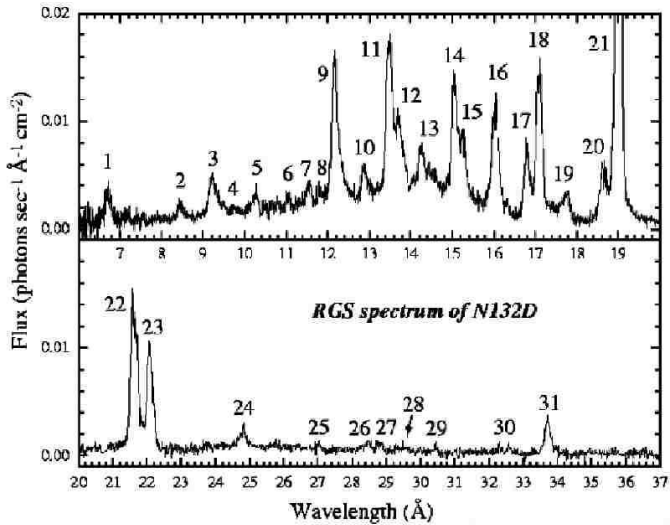,width=7cm,angle=-90,clip=}}
\caption{XMM-Newton RGS spectrum of the SNR N132D in the 
 Large Magellanic Cloud (Behar et al., 2001).
\label{image}}
\end{figure}

\begin{figure}
\centerline{\psfig{file=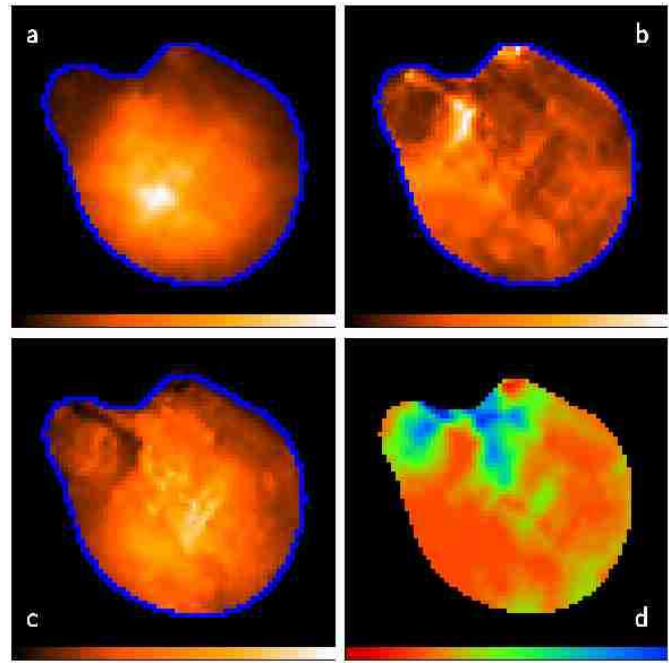,width=8.8cm,angle=0,
clip=}}
\caption{SNR N132D maps of a) the Fe-K line flux at 6.67 keV, b.) the EW of the Fe-K line,
 c.) the differential continuum flux at 6.67 keV interpolated from the 4.4 - 10 keV, 
d.) apparent electron temperature (0.4 - 3.1 keV) derived from just the continuum 
(Behar et al., 2001). 
\label{image}}
\end{figure}


\subsection{Shell-type remnants and the high energy non-thermal spectral tails}

One of the major surprises of SNR X-ray research has been the discovery of high energy tails in the 
spectra of shell-type SNRs. After the discovery of Fe-K line emission in the early 1970's the X-emission 
of shell type remnants has been considered to exclusively arise from shock heated plasma and the radiation 
is thermal, correspondingly. Although even early measurements with HEAO-1 and EXOSAT indicated hard tails 
in young SNRs  
contradicting the straight thermal interpretation the picture has not changed until 1995 when Koyama et al. (1995) 
using ASCA measurements and Willingale et al. (1996) analyzing ROSAT PSPC data discovered that the X-spectra of the 
north-eastern and the south-western caps of the remnant of SN 1006 require a hard tail as well, which is best 
represented by a power law. If the tail is interpreted as synchrotron radiation electrons with energies of 10's of  
TeV must be present. The acceleration to these energies, close to the knee of the cosmic ray spectrum but still 
an order of magnitude or more below, may be provided by diffusive shocks. It seemed that we were close to the solution of 
a problem which has been standing since 1912 when cosmic rays were discovered. SNRs have always 
been considered an important site of cosmic ray 
production. One consequence, however, is that  
 the TeV electrons should also give rise to TeV $\gamma$-rays by inverse Compton scattering off the microwave 
background, and indeed, TeV $\gamma$-rays were subsequently detected (Tanimori et al., 1998). 
Soon after SN 1006 Koyama et al. (1997) added another shell type SNR with a power law spectral tail, i.e. 
RX J1713.7-3946 or G347.3-0.5, which was originally discovered in the ROSAT all-sky survey 
(Pfeffermann \&Aschenbach, 1996). TeV $\gamma$-rays from the north-western quadrant of the remnant 
were reported recently by Enomoto et al. (2002), who could establish a spectrum from 0.4 - 8 TeV. 
They conclude from the TeV spectrum and the upper limit obtained with EGRET on GRCO that the $\gamma$-rays are not 
produced by high energy electrons but result from the decay of $\pi\sp 0$ mesons which are produced 
in collisions of high energy SNR cosmic rays with 
nucleons of an adjacent molecular cloud. A revision of the EGRET data which 
turns the upper limit into a definite flux and a spectrum (Reimer et al., 2002) seems to not exclude 
electron processes.     
  
From the X-ray point of view these remnants are difficult to analyze because the power law contribution  
is a minor component of the overall spectrum which is still dominated by thermal radiation. 
This is different for instance for the supernova remnant RX J0852-4622 or G266.2-1.2 located in the south-east 
corner of the Vela SNR (Aschenbach 1998, Fig. 11). 

\begin{figure}
\centerline{\psfig{file=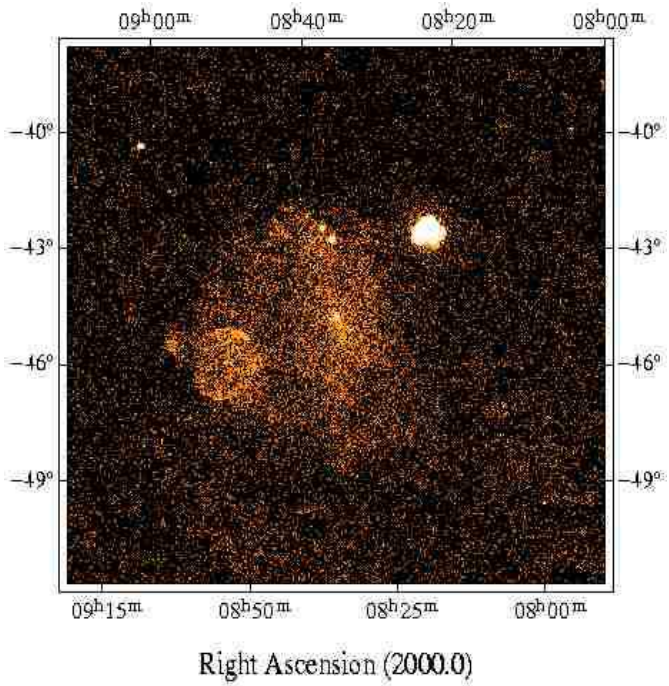,width=8.8cm,angle=0,
clip=}}
\caption{ROSAT all-sky survey image (E$>$1.3 keV) of the Vela SNR, which shows the 
2$\sp{\circ}$ diameter remnant 
RX J0852-4622 in the south-east (Aschenbach 1998).
\label{image}}
\end{figure}

The analysis of the ASCA data shows that the spectrum 
of this shell-type remnant can be represented by a power law without any thermal 'contamination' 
(Tsunemi et al., 2000, Slane et al., 2001). 
This is most evident for the north-western section of the shell whereas there may be contributions to the 
southern spectrum from 
a thermal component, which, however, could be due to radiation from the Vela SNR. 
The observations with XMM-Newton have revealed for the first time the topography of the non-thermal shell. 
The XMM-Newton EPIC pn-camera image is shown in Fig. 12; the section of the shell is broken into a number 
of filaments which are not aligned as well as condensations and indentations. For energies $>$ 1 keV the 
spectrum is a clean power law  with a photon index of 2.6 (Fig. 13, Iyudin et al., 2002); 
an improved fit is obtained with the 'srec' 
model of XSPEC, which takes into account a gradual steepening of the power law towards higher energies 
attributed to electron leakage and/or synchrotron losses. The fit with a power law requires an 
interstellar absorption column density to RX J0852-4622 which is significantly higher than that 
to the Vela SNR, most likely putting RX J0852-4622 behind the Vela SNR. I stress that this is correct 
in a strict sense only for a power law. So far we don't know the intrinsic shape of the spectrum; in case 
of a synchrotron spectrum we don't know whether the electron spectrum really follows a power law 
spectrum.

\begin{figure}
\centerline{\psfig{file=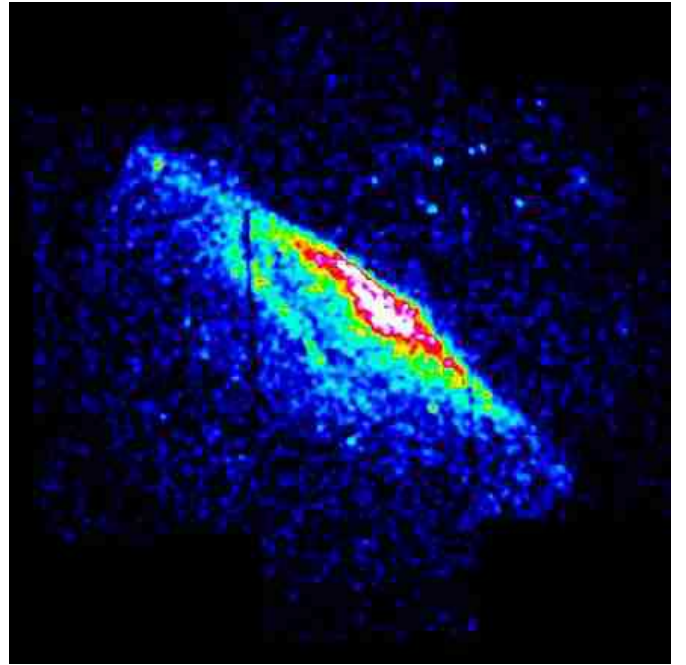,width=8.8cm,clip=}}
\caption{XMM-Newton EPIC-MOS1 image of the north-section section of the SNR RX J0852-4622. 
At low energies, in particular at and around the O lines, there is still thermal emission, which 
mostly likely is from the Vela SNR rather than RX J0852-4622 (Iyudin et al., 2002). 
\label{image}}
\end{figure}

\begin{figure}
\centerline{\psfig{file=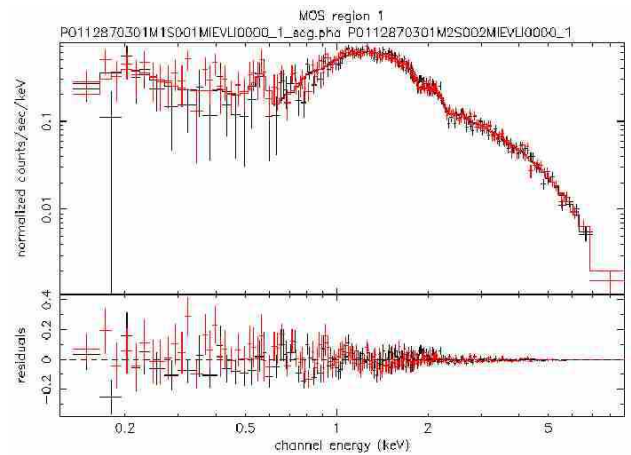,width=6cm,angle=-90}}
\caption{Spectrum of RX J0852-4622 obtained with the XMM-Newton EPIC-MOS1 camera; the fit shows a 
power law modified by interstellar absorption (Iyudin et al., 2002).
\label{image}}
\end{figure}

The only emission line feature in the spectrum is seen at 4.24 keV (Iyudin et al., 2002) assuming 
an underlying power law continuum, although at a fairly low significance level. But such a feature 
was already present in the ASCA spectrum (Tsunemi et al., 2000) although at a slightly lower energy (4.1 keV). 
This line may result from the $\sp{44}$Ti electron capture decay, which produces a vacation in the 
K-shell of the daughter nucleus of  $\sp{44}$Ca or the line is a blend of lines emitted by Ca and Ti. 
The line flux observed is consistent with a relatively young 
and close-by SNR (Aschenbach et al., 1999), and the flux is consistent with the 
1.8 MeV $\gamma$-ray line flux of 
radioactive $\sp{44}$Ti reported by Iyudin et al. (1998). 

Close to the center of RX J0852-4622 a point-like X-ray source  has been found in the ROSAT 
all-sky survey which was suggested as the neutron star remnant of  RX J0852-4622 (Aschenbach 1998). 
But limited by the large RASS error box which contained three optical sources 
no identification was possible. This changed when Chandra observed the region and one source 
now designated CXOU J085201.4-46175 could be singled out (Pavlov et al., 2001). Because of the 
absence of any radio and optical counterpart down to very faint magnitudes and because 
of the spectrum, which looks like a blackbody with kT $\approx$400 eV and an emitting surface of 
 280 m radius for a distance of 1 kpc the source is very likely a neutron star (Kargaltsev et al. 2002). 
No significant pulsation in the 0.001 - 100 Hz range have been found. 
XMM-Newton has observed the central region of  RX J0852-4622 as well confirming the absence any other 
significant X-ray source within a circle of 5 arcmin radius around CXOU J085201.4-46175. Details of the 
results and the spectrum are presented in these proceedings by W.~Becker (2002).

The origin of the high energy X-ray spectral tails is still being discussed controversially, and 
many more SNRs need to be found to solve the puzzle. Remnants dominated by non-thermal rather than thermal 
emission in the X-ray band should have a value as large as possible for the ratio of 
n$\sb{\rm{TeV}}\cdot\rm B\sp{(\gamma + 1)/2}$/$\rm n\sb{\rm{therm}}\sp 2$, 
where n$\sb{\rm{TeV}}$ is the density of relativistic 
electrons in the TeV range, $\rm n\sb{\rm{therm}}$ is the thermal electron density, $\gamma$ is the 
index of the electron power law distribution, B is the perpendicular component of the magnetic 
field in which the electrons radiate synchrotron radiation. If there is a coupling between 
n$\sb{\rm{TeV}}$ and $\rm n\sb{\rm{therm}}$, which might depend on the age of the SNR,  
non-thermal emission is favored in a low density environment. Since the radio emission 
is proportional to n$\sb{\rm{GeV}}\cdot\rm B\sp{(\gamma + 1)/2}$, with not necessarily the same 
$\gamma$, low surface brightness $\Sigma$ radio remnants are preferred candidates to look for 
X-ray synchrotron tails, i.e. remnants which show up at the lower end of the $\Sigma$-D relation. 
This is true for SN 1006, RX J0852-4622, RX J1713.7-3946, RX J0459.1+5147 (Pfeffermann et al., 1991), 
which all demonstrate most clearly the presence of non-thermal X-ray spectra. 
Interestingly, apart from SN 1006 these remnants were discovered by their X-ray emission, which 
means that an unbiased survey in the radio band needs much more increased sensitivity to detect these 
sources. A survey in the X-ray band is severely constrained by interstellar absorption and only 
a survey at energies above 2 - 3 keV, about 10 times more sensitive than the ROSAT survey at 1 keV, 
would detect such remnants. Another implication is that if the majority of SNRs is, as expected, 
from explosion 
of massive progenitor stars, which have had strong stellar winds sweeping out their environment 
to very low densities, a large fraction of SNRs may have escaped detection both in the 
radio and the X-ray bands. And it is this class of remnants which might be the dominant 
cosmic ray contributor. It would also mean that our estimates of the relative number of SNRs 
containing a neutron star is severely biased because we haven't seen the majority of remnants 
because they are  
actually too faint in general to have been observed so far. 

The shell-type power law remnants which have been found so far have typical radio spectra with
an index between $\sim$ 0.3 - 0.5 for the energy flux and the corresponding X-ray spectra have 
an index
typically between 1.5 and 2.0, and the fluxes between the radio and X-ray bands are such
that the energy output per logarithmic energy interval (the $\nu$~F$\sb{\nu}$ distribution)
peaks between the infrared and the EUV, for most of the sources actually in the optical band.
In order to check on the synchrotron nature of the X-rays it is suggested to assess the level
of any optical continuum radiation and to search for polarization of the
optical radiation. This would be an 'experimentum crucis' to decide on the issue of synchrotron 
radiation up into the X-ray band.

\subsection{The Crab Nebula and the 'Bulls Eye'}

Although indicated already in the early Einstein HRI image and more clearly in
the ROSAT HRI image the detailed structure of the Crab Nebula was revealed
only in the superb Chandra images (Fig. 14, Weisskopf et al., 2000). At least three emitting
regions can be discerned, which are the torus, the jet and a counter-jet
and a more diffuse large scale region, extending far to the north-east, for
instance. In particular the region around the pulsar and the main torus
exhibit substantial ellipse-like substructure. 
The existence of the torus-like emission region as an equatorially confined 
pulsar wind nebula was suggested already in the pre-X-ray telescope era in 1975 by 
Aschenbach \& Brinkmann (1975), interpreting lunar occultation 
measurements. In 1985 the existence of the polar outflows (jets) was proposed in a paper by Brinkmann 
et al. (1985) based on a 2-dim Fourier analysis of the {\it{Einstein}} HRI image which were  
even more evident in the ROSAT HRI image if the analysis is restricted to just the spatial high frequency 
components (Aschenbach 1992). The X-ray images show a brightness asymmetry along the torus such that 
the north-western part is much brighter than the south-eastern section, which is attributed to  
Doppler boosting of the electrons leaving the pulsar acceleration zone and hitting the nebula plasma 
with a speed of $\sim$ one third of the speed of light. Since the inclination of the torus out of  the 
plane of the sky is  $\sim$ 30$\sp{\circ}$ (Aschenbach \& Brinkmann, 1975) with the north-western 
section pointing towards the observer, this section should be Doppler brightened (Pelling et al., 1987). 
It is noted that Aschenbach and Brinkmann (1975) suggested instead that the brightening of the 
north-western section is due to a compression of the magnetic field in the nebula by the motion 
of the pulsar against the nebula, the direction of which is pointing fairly close to the position 
of the maximal X-ray enhancement. Interestingly, the same relation holds for the Vela pulsar and its 
synchrotron nebula, i.e. the proper motion vector of the pulsar points to the region of 
maximum brightness. For the polar outflows such large brightness asymmetry has not been measured, and it is 
definitely significantly lower, which in the Doppler boosting model could be explained by either a geometry 
where the polar outflows are not perpendicular to the equatorial outflow, or an electron spectrum which 
is much flatter than the corresponding spectrum in the equatorial wind, or a electron bulk motion  
velocity along the jets much lower than c/3. The spatially resolved XMM-Newton spectra of the Crab 
show that the spectra of the polar outflow tend to be steeper than that of the torus region. If 
the polar outflows are associated with the rotation axis rather than with the magnetic axis the 
jet electrons move at comparatively low  speed.        
 
In the meantime a few more pulsar wind nebulae have been 
added which show the same structural components, so that highly asymmetric pulsar winds associated 
with the equatorial and polar regions appear to be the rule. 
An even more spectacular example is probably the Vela pulsar and its vicinity revealed by the 
Chandra images which have been discussed at this conference as well.

One of the big puzzles about the Crab is still not solved. Even with the deepest 
XMM-Newton and Chandra observations thermal 
 X-ray emission, which is expected from the shock
heated stellar debris and/or the ambient medium, has still not been found. 
In other cases of originally proclaimed Crab-like remnants like SNR 0540-69.3 deeper 
observations revealed in addition to the pulsar and its synchrotron nebula a faint 
shell of thermal emission (Gotthelf \& Wang, 2000, van der Heyden et al., 2001).
But recently a second true Crab-like remnant, a close cousin of the Crab Nebula according to the authors,  
 has been identified as such  using Chandra (Lu et al., 2002). The remnant G54.1+0.3 has been known for a long 
time as a radio source and it belongs to the class of centrally filled and strongly polarized remnants 
with a flat radio spectrum. X-rays have been detected with {\it{Einstein}} and ROSAT, the images 
of the latter observatory show a centrally peaked emission with some structure indicating a jet-like feature 
(Lu et al., 2001). G54.1+0.3 was then resolved with Chandra observations and spectra of various regions 
could be obtained (Lu et al. 2002). The Candra image is shown in Fig. 15. The remnant is 
$\sim$ 2.2' in size at an estimated distance of $\sim$ 5 kpc. It contains a central bright 
point-like source, a surrounding ring or torus, bi-polar elongations or jets and low 
surface brightness diffuse emission. The Chandra spectra show that all these components 
have power law spectra with no evidence of a thermal contribution; a thermal shell is absent as well. 
The flattest spectrum with a photon spectral index of 1.1 is associated with the point-like source, 
the torus has a spectrum of slope 1.64 and the outer diffuse regions come close to 2 for the spectral 
slope; except for the central source the spectra of the other components are pretty close 
to those in the Crab Nebula, the central source is flatter by 0.4. 
Spatially resolved X-ray spectra of the Crab are given by Willingale et al. (2001).  
As in the Crab the brightness of the X-ray torus changes with position vector, i.e. the azimuthal direction. 
In the framework of the Doppler boosting model the electron bulk velocity at the interface with the nebula 
amounts to 0.40$\pm$0.12 times the speed of light, which is basically the same as observed for the Crab 
torus (Lu et al., 2002). It would be interesting to measure the proper motion of the pulsar in G54.1+0.3 to 
see whether the proper motion vector points to the region of maximum brightness as it does for the 
Crab and Vela. If so, the Doppler boosting model has to be reviewed because in its current form 
it does not account for the coincidence of proper motion vector and maximum brightness location.

Recent follow-up observations with the Arecibo radio telescope show the central point-like source 
to be a radio pulsar with P = 136 ms and a $\dot{P}$ such that the characteristic age 
$\frac{P}{2~\dot{P}}$ $\sim$ 2900 yrs and the spin-down luminosity $\dot{E}$ = 
 1.2$\times$10$\sp{37}$ erg/s (Camilo et al. 2002). 
Like in the Crab the pulsar in G54.1+0.3 dominates the picture so that the press has called 
this object 'bulls eye' (c.f. Fig. 15 and www1.msfc.nasa.gov/NEWSROOM/news/releases/2002/ ~02-159.html).
Clearly, based on morphological and spectral similarity the Crab Nebula and G54.1+0.3 belong 
to the same class of objects. But this discovery together with the low 
luminosity of the pulsar of just 1 mJy kpc$\sp 2$, which is among the lowest 
for known young pulsars, shows that SNR pulsars can be extremely faint and many of them might be 
present in SNRs but simply have escaped detection so far. New and upgraded radio telescopes 
like the Arecibo are required to find them.

Another similarity between the Crab Nebula and the nebula of G54.1+0.3 is the mean linear expansion speed
defined as the X-ray radius divided by the age. The apparent X-ray expansion rate of the Crab Nebula is
900 km/s versus 550 km/s for G54.1+0.3; they would be identical if G54.1+0.3 were at a distance of
8 kpc, which is not excluded by current data. But there is also a major difference between the Crab and
G54.1+0.3, which is the radio extent. Whereas the radio diameter of G54.1+0.3 
 (Velusamy \& Becker, 1988) is about the X-ray diameter
the radio Crab is is about twice as large as the X-ray Crab. This could imply that the magnetic field
of G54.1+0.3 is significantly less than that in the Crab Nebula, so that synchrotron losses even in the
X-ray band are not important in G54.1+03. On the other hand the steepening of the X-ray spectra with
increasing distance from the central pulsar, which for the
Crab has usually been attributed to synchrotron losses, is observed in G54.1 +0.3 the same way and 
level.  
Concerning the magnitude of the magnetic field strength a simple estimate shows that the nebular magnetic fields 
are not that much different. The nebular magnetic field scales with the age divided by the pulsar period 
and divided by the linear extent of the nebula. If I take the numbers given above the mean magnetic field 
of G54.1+0.3 should not differ from the Crab Nebula magnetic field by more than a factor of two or three. 
This would mean that the difference of the ratio of the radio versus X-ray size for these two 
sources is not related to the mean magnetic field strength. Consequently, if the pulsar nebula wind is dominated by  
electromagnetic losses rather than by particles, the extent of the nebula is not driven by the 
pulsar wind pressure but is controlled by some outer ambient medium, i.e. its particle and magnetic 
pressure. 
  
This observation also  raises the question whether the spectra we see from the radio range up to the 
X-ray domain across both
the Crab and G54.1+0.3 are intrinsic spectra for this class of source, and that we don't see just the propagation
of electrons injected by the pulsar suffering from synchrotron losses. Is there re-acceleration of the
electrons inside the nebulae out to their edges?

The frequency dependence of the size of the Crab Nebula has been a long standing
issue as it should give insight to the release into and the
propagation mechanism of the electrons through the magnetic field. In the Crab Nebula 
the mean synchrotron lifetime of the electrons in the 0.5 mG field
is $\sim$4 years and increases with frequency as $\nu\sp{-0.5}$ and
a frequency dependent extent is expected.
With the EPIC-MOS cameras of XMM-Newton the X-ray spectrum across the nebula
has been spatially resolved (Willingale \& Aschenbach, 2002). The spectra follow power laws everywhere with
a continuous steepening with increasing distance from the pulsar.
Fig. 16 shows the spectra for four selected regions in comparison
with their optical spectra. Region 1, which is the region inside the
torus
close to the pulsar, shows a very small increase of the spectral index from the optical to the
X-rays by $\sim$0.1, demonstrating that synchrotron losses are negligible
and the electrons are young. But in region 4, which is the far north-west
region at a distance of up to $\sim$ 1.8' from the pulsar the spectrum steepens significantly such that the 
X-ray flux is
far below the extrapolation from the optical about a factor of 60 at 1 keV. This decrement would imply 
an age of the outermost electrons of at most 25 years, because every 4 years the energy of the electrons 
drops by a factor of 2 by synchrotron losses when moving away from the pulsar. This is compatible with the large 
distance of 1.8' from the pulsar only if the electrons move at about 1/7 of the speed of light through the 
nebula. As discussed above, based on the similarity with G54.1+0.3, the challenge of the high propagation 
speed would be significantly relaxed if some re-acceleration occurs inside the nebula. The optical filaments 
might be the sites where diffusive shock acceleration could contribute. Dramatic 
synchrotron losses could also be significantly reduced if the outer regions of the nebula are fed not so much by the 
equatorial wind but by the polar winds where jets might transport the electrons at lower loss rates.    

As for the Crab Nebula it would be very useful to try to measure the non-thermal radiation of 
G54.1+0.3 in the optical (narrow band non-thermal flux and/or polarization) and to look for optical filaments.

\begin{figure}
\centerline{\psfig{file=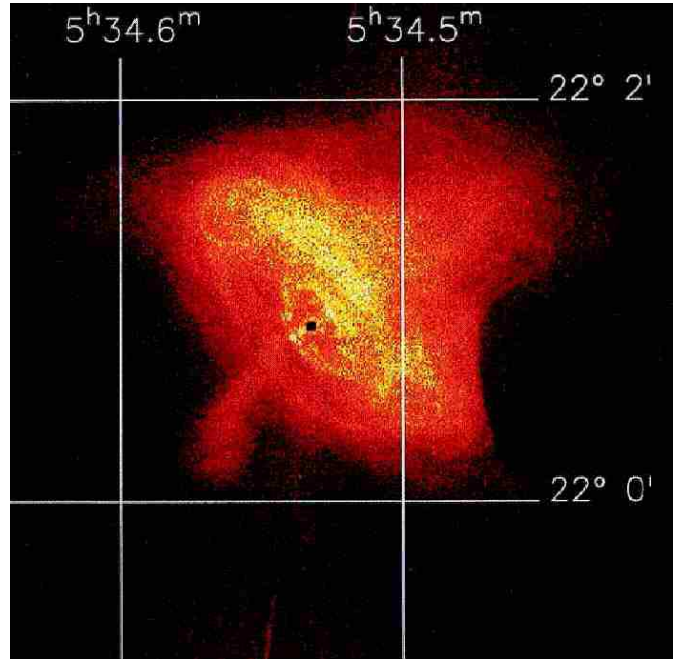,width=8.8cm} }
\caption{Chandra image of the Crab Nebula (Weisskopf et al., 2000).
\label{image}}
\end{figure}

\begin{figure}
\centerline{\psfig{file=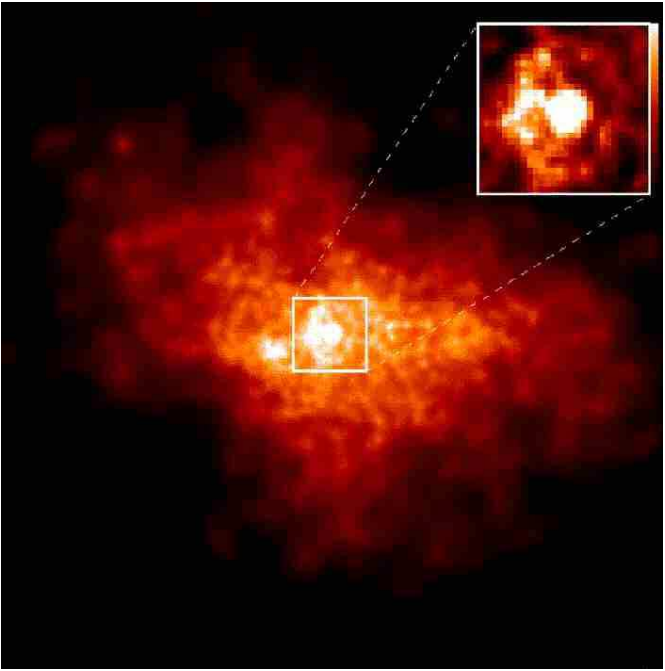,width=8.8cm,clip=} }
\caption{Chandra ACIS-S image of the SNR G54.1+0.3. Size of the image is
127 arcsec square (Lu et al., 2002).
\label{image}}
\end{figure}

\begin{figure}
\centerline{\psfig{file=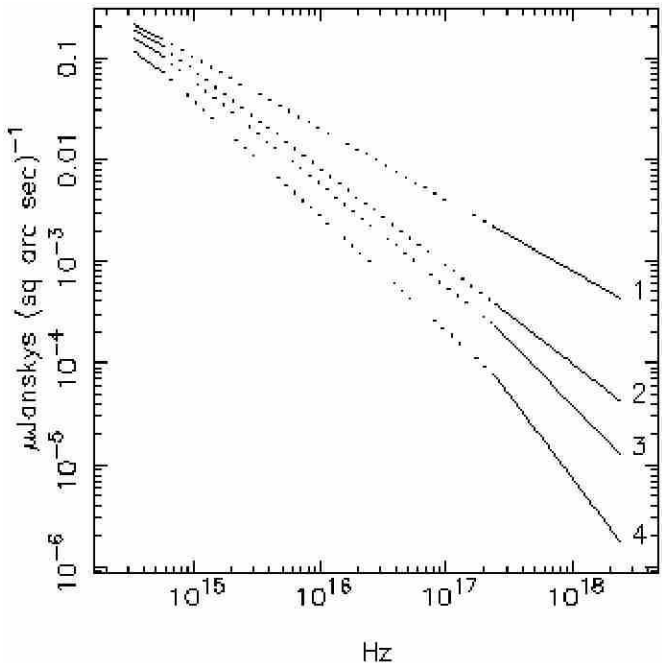,width=8.8cm,clip=} }
\caption{Optical continuum spectra 
and X-ray spectra of the EPIC MOS-cameras from four separate regions
across the Crab Nebula; region 1 corresponds to the inner edge of the torus, region 2 is the 
torus, region 3 is the transition region from the 
torus to the nebula, and region 4 reflects the outskirts of the nebula (Willingale \& Aschenbach 2002).
\label{image}}
\end{figure}


\subsection{SN 1987A}

The explosion of SN 1987A was one of the rare historical astronomical
events and it has triggered research in many astrophysical fields.
In soft X-rays it took a long time, actually almost 3.5 years,
to detect soft X-rays. The first attempt which was carried out with 
a sounding rocket experiment and an imaging X-ray telescope in August 1987, 
just half a year after the 
explosion, provided an upper limit of  1.5$\times$10$\sp{36}$ erg/s, which 
indicated a very low ambient matter density compliant with a wind of a 
blue supergiant (Aschenbach et al., 1987). Since the launch of ROSAT 
SN 1987A has been monitored
regularly and the monitoring is now being taking over by Chandra and XMM-Newton.
Figure 17 shows the light curve, which can be remarkably well fitted
with a t$\sp 2$ relation up to the end of the ROSAT coverage 
about 4000 days after the explosion. Soft X-rays are expected from the
interaction of the supernova shock with matter left from the wind
of the blue supergiant stage and red supergiant phase of the progenitor inside the ring.
The t$\sp 2$ dependence seems to suggest some cylindrical
rather than spherical symmetry of the interaction, which would have
implications on the radial distribution of the matter density surrounding 
SN 1987A.
The XMM-Newton and the Chandra data points tend to exceed 
the t$\sp 2$ best fit. Whether this is an indication that the
shock wave has already reached and partly penetrated the inner
ring around SN 1987A remains to be seen. The Chandra images have
resolved the supernova indicating that the shock wave has reached
the inner surface or is close to contact. The more than proportional 
increase in flux indicates significant mass concentrations in the ring, 
the emission measure of which are notable (Park et al. 2002).  
In order to convert the count rate of the instruments into flux values
the spectrum obtained with the XMM-Newton EPIC-pn camera (Fig. 18)
has been used except for the Chandra data points. The Chandra data points have been 
extracted from the paper by Park et al. (2002).

The spectral fits and the derivation of element abundances is not unique.
The XMM-Newton data, which cover the energy range from 0.3 - 10 keV can be represented
by an optically thin plasma spectrum the temperature of which is  
$\sim$ 0.36 keV and a second component to fit the spectrum above $\sim$ 3.5 keV. 
For the thermal component both collisional ionization equilibrium (CIE) and 
ionization non-equilibrium (NIE) models have been tested and both provide the same parameters for 
temperature, density and abundances. In the context of the NIE models the plasma appears to be 
in collisional ionization equilibrium.    
The high energy component appears to be non-thermal because of the absence of any emission lines 
and it can be represented by a power law of photon index 2.6. The abundance 
values associated with the thermal component are about half solar for all elements 
except Si, S and Fe. Si and S appear to be a few times solar and for Fe only 
an upper limit of 0.03 times solar could be derived. It is noted that both the 
Fe-L lines and the Fe-K line are missing. The absence of Fe is remarkable because the 
modeling of the early hard X-ray lightcurve of SN 1987 A required a substantial mixing 
of $\sp{56}$Co and Fe from the proto-neutron surface to the upper layers (e.g., Sunyaev et al., 1989). 
Where has all this 
Fe gone? Why hasn't it been heated by the reverse shock? And if there is no  
reverse shock, why are Si and S overabundant, the layers of which should have been 
contaminated by Fe? A significant condensation of Fe in dust grains is excluded by infrared measurements 
at about the same epoch of the XMM-Newton measurements. The dust-to-gas ratio in the interaction zone 
is found to be only 0.01\% deduced from ISO measurements (Fischera et al., 2001).
    
Michael et al. (2001) have fit Chandra ACIS CCD spectra with plane-parallel shock models provided in the 
XSPEC S/W package. They can fit the spectrum, which covers the range from 0.35 to 4.5 keV, which just 
one temperature of 2.6 keV, and the abundances of the light elements O, Ne and Mg are significant 
lower than the abundances of Si and S. They are all (except N) lower in absolute terms by a factor of 4 to 5 compared 
with the XMM-Newton values obtained with a CIE model, but the ratio of Si or S over any light element is about 
the same. The relative abundances of the medium Z to the low Z elements are independent of the 
radiation model, the continuum in particular. We have applied the plane-parallel shock model of Michael et al. (2001) 
with exactly the same best fit parameters to the XMM-Newton data but we failed to get an acceptable fit. 
Michael et al. (2001) claim that a high energy component is not needed to fit their Chandra ACIS data, which 
might have to do with the lower high energy cut in the Chandra data compared with the XMM-Newton EPIC pn-data. 
It is not clear what the high energy tails is to be attributed to. Either it is associated with the putative 
pulsar, or it is a high energy bremsstrahlung tail or synchrotron radiation from the shell of SN 1987A, as it 
has appeared in numerous shell-type remnants. Of course, a faint unresolved central synchrotron nebula is not excluded 
at this stage. Much deeper X-ray observations of SN 1987A with XMM-Newton, both with the EPIC cameras and the 
reflection grating spectrometer RGS are needed.

\begin{figure}
\centerline{\psfig{file=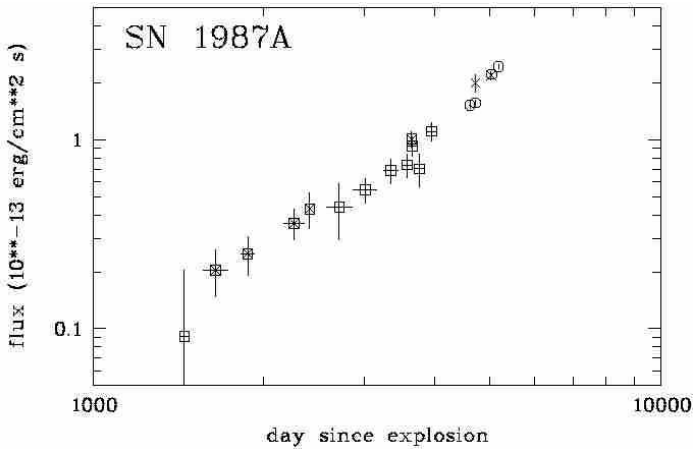,width=6cm,angle=-90}}
\caption{0.5 -2.0 keV lightcurve of SN 1987A
compiled from ROSAT, Chandra (hexagons, from Park et al., 2002) and 
XMM-Newton (crosses) measurements.
\label{image}}
\end{figure}

\begin{figure}
\centerline{\psfig{file=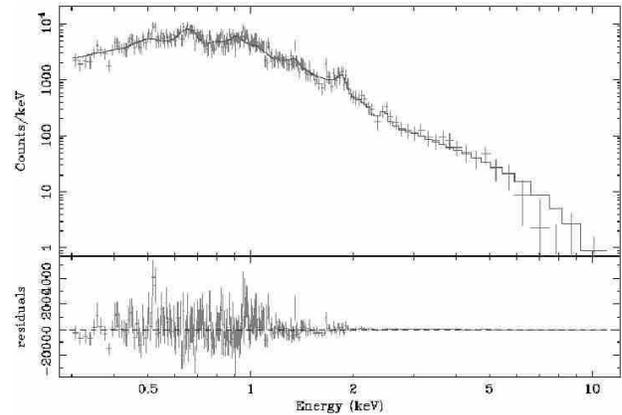,width=5.5cm,angle=-90,clip=} }
\caption{X-ray spectrum of SN 1987A obtained with the EPIC pn-camera.
\label{image}}
\end{figure}

\vskip 0.4cm

\begin{acknowledgements}
I thank the Heraeus foundation for financial support.
\end{acknowledgements}


\clearpage

\end{document}